\documentclass[mnsc,nonblindrev]{informs3}

\OneAndAHalfSpacedXI 

\usepackage{natbib}
 \bibpunct[, ]{(}{)}{,}{a}{}{,}%
 \def\bibfont{\small}%
\TheoremsNumberedThrough     
\ECRepeatTheorems

\EquationsNumberedThrough    

\MANUSCRIPTNO{}

\begin{document}

\MONTH{October}
\YEAR{2023}

\RUNAUTHOR{Roemheld and Rao}

\RUNTITLE{Interference Produces False-Positive Pricing Experiments}


\TITLE{Interference Produces False-Positive Pricing Experiments}

\ARTICLEAUTHORS{%
\AUTHOR{Lars Roemheld, Justin Rao}
\AFF{Zalando SE, \EMAIL{lars.roemheld@zalando.de}} 
} 

\ABSTRACT{%
It is standard practice in online retail to run pricing experiments by randomizing at the \textit{article}-level, i.e. by changing prices of different products to identify treatment effects. Due to customers' cross-price substitution behavior, such experiments suffer from interference bias: the observed difference between treatment groups in the experiment is typically significantly larger than the global effect that could be expected after a roll-out decision of the tested pricing policy. We show in simulations that such bias can be as large as 100\%, and report experimental data implying bias of similar magnitude. Finally, we discuss approaches for de-biased pricing experiments, suggesting observational methods as a potentially attractive alternative to clustering.
}%



\maketitle

%


\section{Introduction}

Online experimentation has become the method of choice for principled decision-making in industry applications, most typically in the form of randomized controlled trials, or ``A/B tests''. However, interference between treatment units can violate the stable unit treatment value assumption (SUTVA) underlying experimentation theory, rendering the measurements obtained through A/B tests biased.

Such bias challenges A/B tests as a “gold standard” for real-world, finite-sample decision making in fields as diverse as pricing (\cite{cooprider2023}), marketplaces (\cite{bajari2023}), and social networks (\cite{karrer2021}). Here, we focus on the first: In online retail pricing experiments, it is standard practice to rely on \textit{article}-level randomization. In such experiments, the retailer's offering of articles is divided into treatment- and control groups such that similar articles experience prices from different pricing policies during an experiment period. This approach is chosen due to legal and PR concerns around price discrimination at the user-level. 

\begin{figure}[b]
\begin{center}
\includegraphics[width=0.9\linewidth]{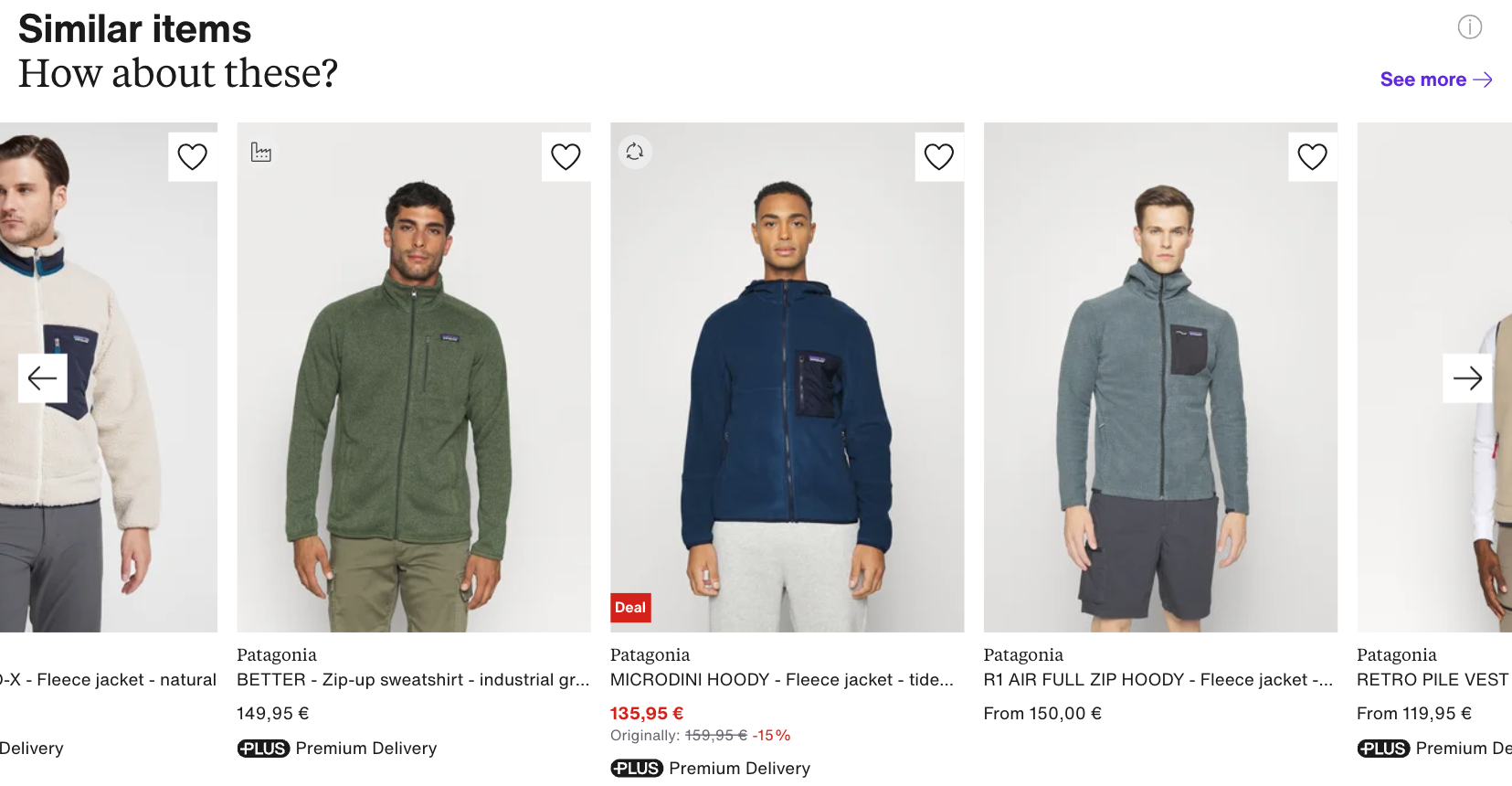}
\caption{Typical display of articles in an e-commerce shop. Standard experiments randomize pricing policies at the article-level to identify their effects on commercial metrics. This screenshot also shows how standard recommendation systems make cross-article prices salient for consumer consideration, contributing to experiment interference through substitution.} \label{recommendation}
\end{center}
\end{figure}

Article-level randomization provides a magnifying glass for practical interference issues in online marketplaces, because the mechanisms for cross-price substitution across similar articles are well-understood. See figure~\ref{recommendation} for a visual demonstration. In this setting, demand patterns of substitution create interference bias between the experimental groups, since the difference between treated and control units is not a good predictor of outcomes following a roll-out decision to 100\% of products (\cite{eckles2016}).
	
In pricing experiments, substitution usually dominates complementarity, leading to inflated differences between the treated and control groups, yielding positive bias and inflated experiment results. This may lead to inflated false discovery rates in A/B tests (\cite{berman2023}), because assumptions about measurement variance will typically assume no bias: bootstrapped standard deviations are usually calculated on so-called A/A test data, where no real intervention was present, leading to the absence of interference in the estimates of standard deviation used for hypothesis testing. Critically, financial expectations of the impact of price cuts may be substantially overstated, leading to misinvestments at the firm level.

\section{Results}

In this working paper, we discuss a practical view of interference in pricing experiments, concluding that naive approaches lead to significantly overstated effects and false positive A/B tests before outlining our plans for further study.

\subsection{Sizing Interference from Substitution}
Interference in pricing experiments mostly comes from substitution (to the inside good), or cross-price elasticity. If the experimenter knew the full-rank elasticity matrix, she wouldn’t need to experiment on pricing; she could know all counterfactuals \textit{a priori}, i.e. all demand reactions to a given price change.

Unfortunately, sparse real-world data usually allows only a coarse approximation of elasticities. To size potential bias from substitution (and complementarity), we use economic primitives to simulate potential interference in an idealized demand system that is consistent with our own e-commerce data and the existing literature. Specifically, we simulate a demand system of $n=10\text{k}$ articles through an elasticity matrix that exhibits three properties:

\begin{enumerate}
\item Standard (negative) own-price elasticity.
\item Substitution focussed on clusters of similar products: every article has a random number of close substitutes.
\item Sharp differentiation between substitution clusters: substitution between two products is never zero, but is moderate for articles not from the same similarity cluster.
\end{enumerate}

\begin{figure}[b]
\begin{center}
\includegraphics[width=0.7\linewidth]{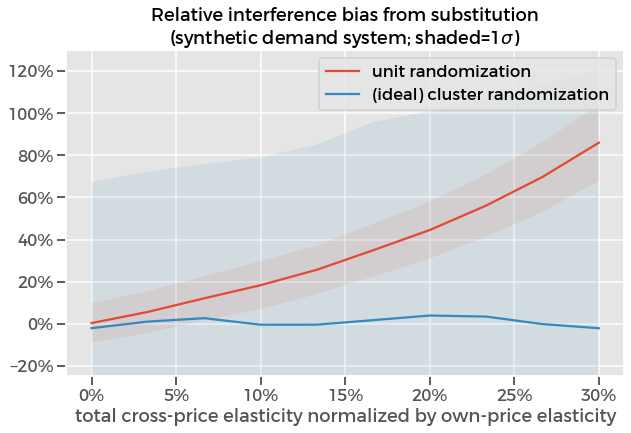}
\caption{Interference bias in simulated pricing experiments. Without any clustered randomization, even moderate cross-price elasticity produces significant interference bias from substitution. Clustered randomization with full knowledge of ideal clusters reduces bias at a cost to variance: measurements get less precise.} \label{syntheticbias}
\end{center}
\end{figure}

With these properties, the elasticity matrix is near-block-diagonal with all negative own-price elasticity on the diagonal and all positive other-price elasticities on the off-diagonals. In this simulated demand system, we know all counterfactuals by design. Consequently, we can compare the estimates of various hypothetical experiments with the real potential outcomes. To this end we permute $p=1000$ different experiment assignments for varying cluster substitution strengths. In figure \ref{syntheticbias} we show that naive article-level randomization leads to heavily biased estimates.

\subsection{Efficient Clustering}

If perfectly separable clusters of articles exist, \textit{and} are known to the experimenter at randomization time, bias can be eliminated entirely, at a cost to measurement variance as shown in figure \ref{syntheticbias}. Under more realistic assumptions, however, perfectly differentiated clusters of substitution may not exist at all,\footnote{In fact, online platforms' endeavors to help customers discover new trends may create budget-driven substitution far beyond the immediately obvious: in figure~\ref{recommendation}, for example, a budget-constrained customer may decide to buy a good deal on sunglasses instead of a fleece jacket, if such a deal becomes salient through recommendation systems.} or may simply not be known to the experimenter. This leaves the experimenter in an unsatisfactory position: Clustering methods can reduce interference bias, but without untestable assumptions about interference-free clusters, they cannot eliminate it, and they require significant additional statistical power. 

\begin{figure}[b]
\begin{center}
\includegraphics[width=0.7\linewidth]{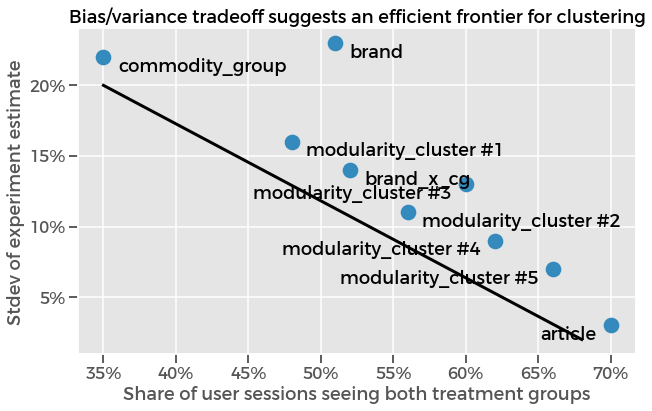}
\caption{Different clustering methods allow choosing appropriate interference- and precision levels. However, using real experiment data from a top EU online retailer, we show that even coarse clusters keep 1 in 3 customers exposed to both treatment groups: Bias can be reduced, but not eliminated.} \label{clusterfrontier}
\end{center}
\end{figure}

In practice, the experimenter thus wants to choose her clustering methods efficiently. Customers' browsing data ("clickstream") can be used to construct a graph of articles often considered in the same session. \textit{Modularity clustering} (\cite{brandes2008}) on this graph can identify subgraphs that are relatively pure (\cite{cooprider2023, karrer2021}): many customers look only at articles from a single cluster, preventing interference between them. In this framework, larger clusters can be purer, but cluster size also increases variance of estimates. This produces a clear trade-off, as shown in figure \ref{clusterfrontier}.

\subsection{Meta-Experiments}

While heuristics such as share of users exposured to both treatment and control groups can help describe the effect of clustering, and make plausible a reduction in interference as in figure~\ref{clusterfrontier}, they do not quantify bias per se. Meta-experiments which contrast experiment arms using different randomization methods quantify bias reduction more directly (\cite{holtz2020}).

In table \ref{tab:metaexp}, we compare effect estimates from the same treatment in 2 real experiments, run sequentially to preserve statistical power. The difference between clustered and article-based randomization shows strong substitution effects on a similar order of magnitude as our simulation results would suggest (see figure \ref{syntheticbias}), leading to estimates that differ by roughly $6\sigma$.

\begin{table}
\begin{center}
\begin{tabular}{l r r r} 
 & effect (clustered) & effect (article-based) & relative bias \\
 \hline
country 1 & 41\% (±5\%-pts) &  61\% & +50\% \\ 
country 2 & 35\% (±8\%-pts) &  62\% & +77\% \\
\end{tabular}
\caption{Commercial treatment effect estimates for the same treatment from 2 separate experiments using clustered and article-based randomization.} \label{tab:metaexp}
\end{center}
\end{table}

\section{Conclusion and Outlook}

Our results suggest that interference bias from substitution behavior in naive pricing experimentation is significant. While clustering methods can reduce such bias, they cannot eliminate it under realistic assumptions. Most importantly, clustered randomization does not help quantify the remaining amount of bias while losing statistical power and plausible generalizability.

In most settings, interference bias will be positive, leading to false positive A/B-test results. This finding suggests that some A/B-tested algorithmic pricing ``improvements'' may fail to materialize in reality.



In light of the untestable assumptions required to trust the results of (clustered) pricing A/B tests, experiments and observational methods of causal inference may be viewed as lying on the same bias/variance trade-off, rather than following a strict epistemic hierarchy. As cluster sizes grow larger on the frontier sketched in figure \ref{clusterfrontier}, interference reduction becomes more plausible, but positivity of treatment assignment becomes more questionable along with statistical power losses. This observation suggests that observational methods of causal inference may be meaningful alternatives for the experimenter to conduct controlled, but not necessarily randomized experiments.

In further work, we will compare the results of observational and experimental effect estimates, relying on modern synthetic control methods. Since online experimentation settings often offer rich data on potential experimental units that were excluded from experimentation, such as other markets, timeframes, or product categories, we believe that such non-randomized data can plausibly be used to construct meaningful counterfactuals.

Such methods introduce untestable identifying assumptions of their own. However, in data-rich environments of online experimentation, these assumptions can often be ``backtested'' relatively well. We argue that seen from a bias/variance perspective, observational methods should at least be a promising secondary analysis for real-world decision making using online experimentation.

\bibliographystyle{nonumber}

\end{document}